\documentclass[journal=jacsat,manuscript=article]{achemso}

\usepackage{chemformula} 
\usepackage[T1]{fontenc} 
\usepackage{amssymb}

\author{Francesco Macheda}
 \email{francesco.macheda@kcl.ac.uk}
 \affiliation{%
Department of Physics, King's College London, Strand, London WC2R 2LS, United Kingdom
}%
 \author{Samuel Ponc\'e}%
 \email{samuel.ponce@epfl.ch}
 \affiliation{%
Department of Materials, University of Oxford, Parks Road, Oxford OX1 3PH, United Kingdom}
 \affiliation{%
Theory and Simulation of Materials (THEOS), École Polytechnique Fédérale de Lausanne, CH-1015 Lausanne, Switzerland
}%
 \author{Feliciano Giustino}%
 \email{fgiustino@oden.utexas.edu}
 \affiliation{%
Oden Institute for Computational Engineering and Sciences, The University of Texas at Austin, Austin, Texas 78712, USA }
 \affiliation{%
Department of Physics, The University of Texas at Austin, Austin, Texas 78712, USA
}%
\author{Nicola Bonini}%
 \email{nicola.bonini@kcl.ac.uk}
\affiliation{%
Department of Physics, King's College London, Strand, London WC2R 2LS, United Kingdom
}%

\title{Theory and computation of Hall scattering factor in graphene}

\begin{document}


\begin{abstract}
The Hall scattering factor, $r$, is a key quantity for establishing carrier concentration and drift mobility from Hall measurements; in experiments it is usually assumed to be 1. In this paper we use a combination of analytical and \textit{ab initio} modelling to determine $r$ in graphene.
While at high carrier densities $r \approx 1$ in a wide temperature range, at low doping the temperature dependence of $r$ is very strong with values as high as 4 below 300~K. These high values are due to the linear bands around the Dirac cone and the carrier scattering rates due to acoustic phonons. At higher temperatures $r$ can instead become as low as $0.5$ due to the contribution of both holes and electrons and the role of optical phonons. Finally, we provide a simple analytical model to compute accurately $r$ in graphene in a wide range of temperatures and carrier densities.
\end{abstract}

Since the discovery of its exceptional room-temperature mobility \cite{Novoselov666, Zhang2005}, graphene has attracted unprecedented interest and research effort to understand its superior electrical properties and exploit them for advanced technological applications.
At the heart of many of these studies are the Hall measurements of this material, which provide the access to key  physical information of semiclassical~\cite{BOLOTIN2008351} and purely quantum nature~\cite{Novoselov666, Zhang2005}.
In particular, these measurements are routinely used to estimate the carrier concentration and the drift mobility of samples.
However, the accurate determination of these quantities relies on the knowledge of the Hall scattering factor $r$. Indeed, the drift mobility is the ratio between the Hall mobility (that comes directly from the Hall measurement) and $r$, and the carrier concentration is directly proportional to $r$ \cite{Popovic}. 
This factor is commonly assumed to be around unity (see, for instance, Refs.~\citenum{Novoselov666,BOLOTIN2008351}), but this assumption is implicitly based on studies done on bulk semiconductors with quasi-parabolic band dispersion, where $r$ generally shows a weak dependence on temperature and scattering mechanisms.
Graphene instead is a two-dimensional (2D) material characterised by a linear dispersion of the electronic bands, and it displays non-trivial behaviours of the energy dependence of the carrier scattering rates in different temperature and doping regimes. 
There is therefore a clear degree of uncertainty about the value of $r$ in graphene; as a consequence, extracting accurately from experiments the carrier density and the drift mobility is not straightforward and requires particular care.
This is also the case in top-gated graphene where the charge density on the sample can be assessed through the gate capacitance, a parameter that however is often determined via a Hall measurement~\cite{doi:10.1021/nl103306a, doi:10.1021/nl201358y,Brown2019,C7NR06330J}.
Interestingly, the need of a detailed study of the Hall scattering factor is of high relevance not only in graphene but more in general in 2D materials~\cite{doi:10.1021/nn402377g, Ma_2015}.

In this paper, using analytical and {\it ab initio} computational modelling, we show that in graphene the intrinsic $r$ follows non-trivial trends as a function of carrier concentration and temperature, with values that can be far greater or smaller than unity.
This clearly can lead, for instance, to very large differences---even up to a factor 4---between the carrier density routinely estimated in a Hall measurement and the real charge concentration of the sample.

For low doping concentrations and temperatures below $\sim$ 300~K, we evaluate $r$ using an analytical model that is deduced starting from the Boltzmann Transport Equation (BTE) in the assumption of perfectly isotropic band dispersion and that depends on the electronic density of states, the Fermi velocity and the carrier scattering rates, which can be evaluated via simplified models.

Above room temperature or at high doping, the validity of the model can be questioned. Indeed, previous works have shown the necessity to exactly solve the BTE in these regimes in order to accurately quantify the resistivity~\cite{PhysRevB.90.125414}. Therefore in these regimes we evaluate $r$ using the full solution of the BTE, obtained by means of the most recent techniques and developments regarding the first-principles calculation of transport properties \cite{PhysRevB.94.085204,PhysRevB.98.201201,PhysRevB.97.121201,Ponc__2020}. 
For this, the electronic and vibrational properties are computed with density functional theory~\cite{PhysRev.140.A1133} and density functional perturbation theory~\cite{RevModPhys.73.515} as implemented in QUANTUM ESPRESSO \cite{Giannozzi_2017}, within the local-density approximation (LDA)~\cite{PhysRevB.45.13244,PhysRevB.23.5048}. The \textit{ab-initio} quantities are interpolated with a Wannier interpolation scheme\cite{RevModPhys.84.1419,MOSTOFI2008685} as implemented in EPW ~\cite{PhysRevB.76.165108,NOFFSINGER20102140,PONCE2016116}. To fully converge the transport observables we use very fine $\mathbf{k/q}$-point grids (up to $2160\times2160\times1$). Convergence tests and further computational details are given in the Supporting information. 
The results presented here are for $p$-doped graphene, but, as shown in the Supporting Information, $n$-doping gives very similar results. Flexural phonons are neglected in this work. This is because in free-standing graphene the coupling between flexural phonons and electrons near the Dirac cone is of second order by symmetry \cite{PhysRevB.76.045430}. In addition, as shown in Ref. \cite{PhysRevB.96.075448} ~, even in gated graphene the coupling  has negligible effects on transport properties.
Very interestingly, this careful {\it ab initio} study of $r$ shows that the analytical model developed for low doping and $T \lesssim 300K$ can reproduce with a very good accuracy the values of $r$ in a wide range of carrier concentrations and temperatures.

To introduce the analytical model for $r$ we start from the linearized BTE ~\cite{PhysRevB.94.085204,PhysRevB.98.201201,Ponc__2020}:
\begin{align}
&-ev_{n\mathbf{k}}^{\beta}\frac{\partial f^0_{n\mathbf{k}}}{\partial \epsilon_{n\mathbf{k}}}-\frac{e}{\hbar}(\mathbf{v}_{n\mathbf{k}}\wedge \mathbf{B})\cdot\nabla_{\mathbf{k}}\partial_{E^{\beta}}f_{n\mathbf{k}}=\sum_{m}\nonumber\\
&\int_{BZ}\frac{d^3q}{\Omega_{BZ}}[\tau^{-1}_{m\mathbf{k+q}\rightarrow n\mathbf{k}}\partial_{E^{\beta}}f_{m\mathbf{k+q}}-\tau^{-1}_{n\mathbf{k}\rightarrow m\mathbf{k+q}}\partial_{E^{\beta}}f_{n\mathbf{k}}]
\label{eq:BTE}
\end{align}
where $\{n\mathbf{k}\}$ are the band and $\mathbf{k}$-point indexes, $\epsilon_{n\mathbf{k}}$ and $\mathbf{v}_{n\mathbf{k}}$ are the electron band energy and velocity, $f_{n\mathbf{k}}$ is the out of equilibrium electronic population, $\mathbf{B}$ and $E^{\beta}$ are the magnetic field and the electric field along the $\beta$ cartesian direction and $\tau^{-1}_{m\mathbf{k+q}\rightarrow n\mathbf{k}}$ is the partial scattering rate as defined in Ref.~\citenum{Ponc__2020}. 
Under the assumption of perfectly isotropic bands energies and a quasielastic scattering, the collision term on the right hand side of Eq.~\ref{eq:BTE} can be written {\it exactly} as $1/\tau^{0}_{n\mathbf{k}}\partial_{E^{\beta}} f_{n\mathbf{k}}$~, where the scattering time includes the factor $(1-\cos \theta)$ as discussed in Ref. ~\citenum{PhysRevB.45.3612} . In this case $1/\tau^{0}_{n\mathbf{k}}$ is entirely due to the scattering with acoustic phonons and can then be calculated as in Refs.~\citenum{PhysRevB.45.3612,PhysRevB.90.125414}.
In passing by, it is important to point out that the right hand side of Eq.~\ref{eq:BTE} can be written in the form $1/\tau^{0}_{n\mathbf{k}}\partial_{E^{\beta}} f_{n\mathbf{k}}$ also when the out-of-diagonal components of the collision term of the BTE are neglected. In this case, the approach is an {\it approximation} (known as SERTA approximation, see Supporting Information) and $\tau^{0}_{n\mathbf{k}}$ is the quasiparticle lifetime (see Ref.~\citenum{Ponc__2020}). To build the analytical model we now write the BTE equation in cylindrical coordinates,
centered around the $\mathbf{K}$-point, using the linear band dispersion of graphene (valid in the energy window of $1.5$ eV around the Dirac cone \cite{PhysRevB.76.205411}). Taking the magnetic field along the $z$ direction (perpendicular to the graphene plane) we get (see Supporting Information) $\partial_{E^{\rho}}f_{n\mathbf{k}}=(-1)^n ev_F\frac{\partial f^0_{n\mathbf{k}}}{\partial \epsilon_{n\mathbf{k}}}\tau^{0}_{n\mathbf{k}}$ and $
\partial_{E^{\theta}}f_{n\mathbf{k}}=-\frac{e^2}{\hbar}v^2_FB^z\frac{1}{k^{\rho}}\frac{\partial f^0_{n\mathbf{k}}}{\partial \epsilon_{n\mathbf{k}}}(\tau^{0}_{n\mathbf{k}})^2$ where $v_F$ is the Fermi velocity and the $\mathbf{k}$ dependence is through the modulus $k^{\rho}$ only. Inserting these expressions in the definition of the conductivity tensor $\sigma_{ij}=\sum_n \frac{2}{\Omega \Omega_{BZ}}\int_{BZ} r dr d\theta \frac{\partial f_{n\mathbf{k}}}{\partial \epsilon_{n\mathbf{k}}} v^i_{n\mathbf{k}}\partial_{E^j}f_{n\mathbf{k}}$ where $i,j$ are cartesian indexes, $\Omega$ and $\Omega_{BZ}$ are the direct and reciprocal space areas and using the definition of $r$ and the transformation between cartesian and cilindrical coordinates we find:
\begin{align}
r=\frac{n_e |e|}{B_z}&\frac{\sigma_{12}}{\sigma^2_{11}} = -2\pi v^2_F \hbar^2 n_e \times \nonumber \\
&\times \frac{\sum_n(-1)^n \int_{-\infty}^{\infty}d \epsilon \frac{\partial f^0_{n}}{\partial \epsilon} (\tau^{0}_{n})^2 }{\left( \sum_n (-1)^n \int_{-\infty}^{\infty} d\epsilon \epsilon\frac{\partial f^0_{n}}{\partial \epsilon} \tau^{0}_{n}   \right) ^2}
\label{eq:rdef}
\end{align}
where $n_e$ is the carrier density, $|e|$ the modulus of the electric charge, the sum over $n$ is performed on the $\pi$ and $\pi^*$ bands and we have replaced the $\mathbf{k}$-dependence in favour of the energy variable $\epsilon$, as measured from the Dirac cone energy $E_D$, using the band energy dispersion relation. 

Using Eq.~\ref{eq:rdef} we can predict the behaviour of $r$ in different temperature ranges by considering the energy and temperature scaling of the scattering times $\tau^{0}_n(\epsilon)$. 
In the  \textit{Bloch-Gruneisen (BG) regime} ($T<T_{BG}=2\hbar v_{TA/LA} k_F/k_B$ where $v_{TA/LA}$ is the transverse/longitudinal acoustic sound velocity and $k_F$ is the Fermi cystal quasi-momentum) a close expression of $\tau^{0}_{n}$ as a function of energy is non-trivial \cite{PhysRevB.90.125414,C9CP05740D}, but we can evaluate it at the Fermi level $E_F$ as $\tau_n^{0}(E_F)\propto \frac{(k_B T)^4}{E_F}$ \cite{PhysRevB.77.115449}; all the other quantities in Eq.~\ref{eq:rdef} can be evaluated at $\epsilon=E_F$ as $\frac{\partial f^0_{n}}{\partial \epsilon} \approx -\delta(\epsilon-E_F)$ . With this approximation we have that $r$ is a constant as $r\propto n_e/E_F^2$ and $n_e \propto \pm \int_{E_F}^{E_D=0} E dE=\pm E_F^2/2$. Thus, in this regime we expect $r$ to be weakly dependent on temperature and doping.
In the \textit{equipartition (EP) regime} ($T_{BG}<T\lesssim 270$~K, where the upper bound indicates the temperature at which the population of the $A_1'$ phonon mode becomes non-negligible~\cite{PhysRevB.90.125414}) it holds that $\hbar \omega_{\mathbf{q},TA/LA} << k_B T$, that the electron-phonon scattering is quasielastic and that the electronic populations do not change appreciably over a length of $\hbar \omega_{\mathbf{q},TA/LA}$ around the Fermi level; using this information the expression for the inverse scattering time is found to scale as $1/\tau^{0}_n(\epsilon)\propto |\epsilon|k_BT$. However, note that, at $\epsilon=0$,  
the total scattering rate $1/\tau^{0}_n(\epsilon)$ cannot be exactly zero because of small but non-vanishing contributions due to scattering with optical phonons, higher order scatterings or possible scattering with defects or boundaries.
By taking in account these contributions, the validity of the scaling of $1/\tau^{0}_n(\epsilon)$ is assumed to hold true up to $1$ meV below/above $E_D$ (see Supporting Information).
It is thus clear that in this regime the numerator of Eq.~\ref{eq:rdef} can attain large numerical values while the denominator converges quickly thanks to presence of $\epsilon$ in the integrand. In this regime we thus expect unusually large values of $r$, especially for low carrier concentrations.
\begin{figure}[!t]
\centering
  \includegraphics[width=1\linewidth]{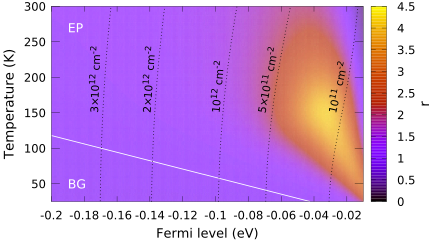}
  \caption{Hall scattering factor $r$ as a function of temperature and Fermi level (corresponding carrier concentrations are indicated with dashed lines) obtained using Eq. \ref{eq:rdef} with $\tau^{0}_{n\mathbf{k}}$ from Refs.~\citenum{PhysRevB.77.115449,PhysRevB.90.125414}. We used a cutoff of $1$~meV around the Dirac cone. The white line separates the Bloch-Gruneisen (BG) and the equipartition (EP) regimes.}
 \label{fig:HallBGEPDFT}
\end{figure}

The scaling considerations discussed above can be verified by inserting the scattering models for acoustic phonons of Refs.~\citenum{PhysRevB.90.125414,PhysRevB.77.115449} inside Eq. \ref{eq:rdef} and studying the BG and EP regimes. The parameters for the scattering are evaluated at a DFT level (see Supporting Information).  In the EP regime we include also inelastic scattering with the $A_1' $ and optical $\Gamma$ phonon modes as modelled in Ref.~\citenum{PhysRevB.90.125414} and using Matthiessen’s rule.
While the inclusion of inelastic terms makes Eq. \ref{eq:rdef} not exact, we expect this approximation to be accurate since the scattering from optical $\Gamma$ phonons is negligible in this regime and the scattering from $A_1' $ phonons is weak (but non-zero) around room temperature.
The predictions of the analytical model are shown in Fig.~\ref{fig:HallBGEPDFT}.In the EP regime, going from high to low doping, $r$ tends to increase up to a maximum value of $\sim 4$ at $\sim 30$~meV below the Dirac cone and $\sim 150$~K (around $1.5 \times 10^{11} cm^{-2}$). At higher temperatures and at low carrier concentrations instead we witness low values of $r$ ($\sim 0.5$). Such low values are due to the fact that here $E_F$ is very near to $E_D$ and the spread of the Fermi-Dirac distribution is large. In this case, both electrons and holes contribute to the value of $r$ but with different signs and thus compensate each other; in other words, the subtraction of the scattering times at the numerator of Eq. \ref{eq:rdef}, which is not hampered by a sharp $\frac{\partial f^0_{n\mathbf{k}}}{\epsilon_{n\mathbf{k}}}$, turns out to be effective.
 These results clearly show that the common assumption of $r=1$ in a Hall experiment performed in the EP regime can lead, for instance, to very rough estimates of the charge concentration in a graphene sample.
The dependence of $r$ upon doping and temperature is instead very mild in the BG regime and its value is around unity when using the DFT parameters. To test the robustness of these results, we have also evaluated Eq.~\ref{eq:rdef} using the {\it ab initio} GW parameters given in Ref.~\citenum{PhysRevB.90.125414} for both the electron-phonon coupling (EPC) parameters and $v_F$; in this case we obtain values for $r$ very similar to the DFT ones if we use the same Fermi level. If instead we compare GW and DFT results at equal carrier concentrations, the Hall scattering factor from GW parameters turns out to be $\sim 30 \% $ larger than the DFT one over the whole range of temperatures.
This is mostly due to the change in $v_F$, not in the EPC parameters. Indeed, if we evaluate Eq. \ref{eq:rdef} at equal carrier densities using the GW EPC parameters but we keep $v_F$ at its DFT value, the Hall scattering factor is very similar to $r^{DFT}$.
This result is very interesting because engineering the Fermi velocity in graphene---a possibility demonstrated in Ref.~\citenum{Hwang2012}---can open up the opportunity to control the value of $r$.

\begin{figure}[!t]
\centering
  \includegraphics[width=1\linewidth]{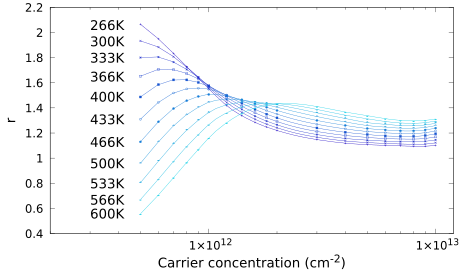}
  \caption{Hall scattering factor $r$ obtained from the full {\it ab initio} solution of the BTE as a function of carrier concentrations, at different temperatures.}
 \label{fig:Hall_hole_density}
\end{figure}

We now proceed to study the high-temperature regime ($ T \gtrsim 270$~K)  for various and non necessarily low carrier concentrations. As mentioned above, in this regime the validity of Eq.~\ref{eq:rdef} and the existence of simplified models for $\tau^{0}_{n\mathbf{k}}$ is not straightforward. In particular, one would expect that the importance of {\it ab initio} calculations in this regime is given by {\it i}) the relevance of inelastic scattering, {\it ii}) the lack of a simple but accurate model for the coupling of optical phonons at $\Gamma$ to electrons (see Supporting Information), {\it iii}) the failure of Matthiesen's rule and {\it iv}) the limited validity of the symmetry-based models for the EPC to $\mathbf{k}$ points very close to the Dirac cone (condition which is satisfied only at very low doping). 
We thus resort to the full {\it ab initio} solution of the BTE equation; the results for $r$ are shown in Fig. \ref{fig:Hall_hole_density}. Here we notice that there is a cross-over in the temperature dependence of $r$ happening for concentrations $10^{12}$cm$^{-2}<n<2\times 10^{12}$cm$^{-2}$. The first regime, at low doping, shows that $r$ decreases with temperature, a trend consistent with the one of Fig. \ref{fig:HallBGEPDFT}. In the second regime ($n>2\times 10^{12}$~cm$^{-2}$) $r$ instead gently increases with temperature and presents a quite smooth doping dependence.

\begin{figure}[h]
\centering
  \includegraphics[width=1\linewidth]{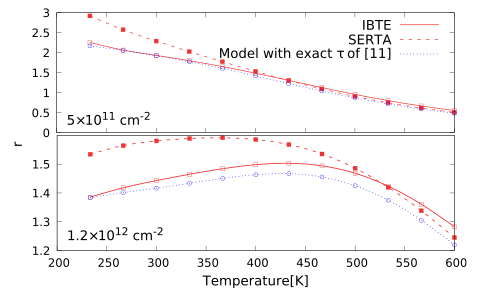}
  \caption{Comparison among the values of $r$ as a function of temperature obtained from the full iterative solution of the BTE (IBTE) (continuous red line), using the SERTA approximation (dashed red line) and from the model discussed in the text (dotted blue line) at a carrier concentration of $5\times10^{11}$ ~cm$^{-2}$ (top panel) and $1.2 \times 10^{12}$~cm$^{-2}$ (bottom panel).} 
  \label{fig:comparison3}
\end{figure}

Surprisingly, the {\it ab initio} values of $r$ (Fig.~\ref{fig:Hall_hole_density}) can be well reproduced in a very wide range of temperatures and carrier concentrations by using the model of Eq.~\ref{eq:rdef} including both the acoustic and optical scattering rates of Ref.~\citenum{PhysRevB.90.125414} with the Matthiesen's Rule. 
The agreement between the model and \textit{ab initio} data is shown in Fig.~\ref{fig:comparison3} where, as an example, we show the behaviour of $r$ at low doping ($5\times10^{11}$ ~cm$^{-2}$) and at a carrier density ($1.2 \times 10^{12}$~cm$^{-2}$) at which $r$ displays the cross-over in the temperature dependence observed in Fig.~\ref{fig:Hall_hole_density}.
It is important to point out that this agreement between the two methods happens despite the inaccuracy of the model~\cite{PhysRevB.90.125414} in describing the resistivity at high temperatures and carrier densities.
The fact that the analytical model turns out to be a good tool to compute $r$ is due to the compensation of scaling effects between the numerator and the denominator in Eq.~\ref{eq:rdef}. In particular, this means that the high-temperature behaviour is well reproduced despite the use of simplified models for the optical phonons scattering from Ref. \cite{PhysRevB.90.125414}.
A similar behaviour is also witnessed in other transport quantities that are expressed as ratios, as for example the Seebeck coefficient (see, for instance, Refs.~\citenum{PhysRevB.94.085204, Rittweger_2017, PhysRevResearch.2.033055} where various approaches are analysed in different materials).

\begin{figure}[!t]
\centering
  \includegraphics[width=1\linewidth]{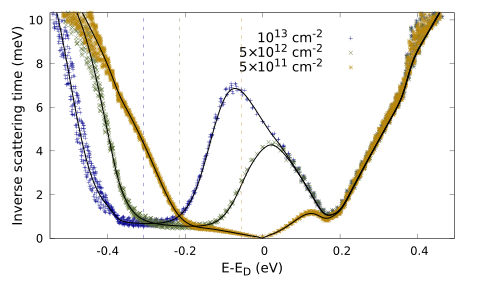}
  \caption{{\it Ab initio} scattering rate for three different doping concentrations ($5 \times 10^{11}$~cm$^{-2}$, $5 \times 10^{12}$~cm$^{-2}$ and $10^{13}$~cm$^{-2}$) at 300K, as a function of the distance from the Dirac cone energy. The black continuous line is the fit of the SERTA scattering time done via the following expression $\tau_n^{0}(\epsilon)= (-1)^n\frac{v_F^2\Omega_{BZ}}{2\pi \epsilon N_\mathbf{k}}\sum_{\mathbf{k}} \delta(\epsilon^{DFT}_{n\mathbf{k}}-\epsilon) \tau^{0,DFT}_{n\mathbf{k}}$ (see Supporting Information), which can be inserted in Eq. \ref{eq:rdef}. The vertical dashed lines represent the Fermi levels for each doping concentration.  }
 \label{fig:invtau}
\end{figure}

In Fig.~\ref{fig:comparison3} we also plot the temperature dependence of $r$ in the SERTA approximation. It is interesting to notice that this approximation is slightly less precise than the analytical model, especially at low temperature, but it still captures the physical trends of the exact solution (in the Supporting Information we show in detail that the trends of the scattering rates of Refs.~\citenum{PhysRevB.45.3612,PhysRevB.77.115449} are in agreement with the SERTA ones). Therefore the SERTA approximation---in which the $\tau^{0}_{n\mathbf{k}}$ are the physically observable quasiparticle lifetimes---can be used to analyse and understand the microscopic mechanisms that are at the hearth of the change in behaviour of $r$ at high carrier concentration.
Indeed, in Fig.~\ref{fig:invtau} we show  the SERTA inverse scattering times for three different carrier concentrations ($5 \times 10^{11}$~cm$^{-2}$, $5 \times 10^{12}$~cm$^{-2}$ and $10^{13}$~cm$^{-2}$) at 300~K.
We notice that the almost linear behaviour of $1/\tau^{0}_{\mathbf{k}}$ around $E_D$---the fingerprint of acoustic phonon scattering---is evident for the lowest concentrations, but tends to disappear with increasing doping. In addition, the mark of optical phonon emission, which is the peak situated at around $200$~meV above $E_D$, moves at lower energies while maintaining its position around $200$~meV above the Fermi level (indicated by vertical dashed lines for each concentration). This means that at high doping the numerator of Eq. \ref{eq:rdef} is not prone to singular behaviours and therefore we expect it to be much less sensitive to the form/nature of the scattering (in general, the denominator of Eq. \ref{eq:rdef}, related to the drift mobility, is in this regime always smoothly behaved). As a result, the spread of the values of the Hall scattering factor as a function of temperature is strongly reduced.

In conclusion, we have determined from first principles the intrinsic behaviour of the Hall scattering factor in graphene for a wide range of carrier densities and temperatures.
Our results show that at high carrier densities the fact that the carriers are far from the Dirac cone and strongly coupled with optical phonons makes $r$ weakly temperature dependent with values around unity. 
Instead, at low doping, this quantity displays a strong temperature dependence with values much larger than unity below room temperature. The high $r$ is due to the nature of the scattering of the carriers close to the Dirac cone with acoustic phonons in conjunction with the two-dimensional conic geometry of the bands. This clearly suggests that in two-dimensional materials with non-parabolic band dispersion the common practice of assuming $r=1$ in Hall measurements should require careful examination.
Our results also show that, even though for other transport quantities a precise solution of the BTE is needed in order to accurately determine the transport coefficient, for the Hall scattering factor the expression of Eq. \ref{eq:rdef} with analytic scattering models is an accurate tool to compute $r$ in a wide range of temperatures and carrier densities. 
Finally, it is worth pointing out that a better understanding of the Hall scatting factor in graphene and a simple tool to calculate it can be of great importance to explore and optimise graphene-based devices. For instance, a direction of particular interest is the current research on graphene Hall sensors, that have the potential to outperform traditional magnetic sensors based on semiconductors~\cite{CHEN2015585,Song2019,Collomb2019,s41467-020-18007-5,Schaefer2020}.

This document is the unedited Author’s version of a Submitted Work that was subsequently accepted for publication in NanoLetters, copyright © American Chemical Society after peer review. To access the final edited and published work see 

https://pubs.acs.org/doi/abs/10.1021/acs.nanolett.0c03874 .
\\
\textbf{Supporting Information---}The Supporting Information contains i) the computational details ii) the convergence tests for the interpolation of the \textit{ab-initio} quantities and the transport observables iii) a discussion of the difference between hole and electron transport iv) the plots of the scattering rates v) a detailed derivation of Eq. \ref{eq:rdef} and vi) the list of DFT and GW parameters in the analytical formula for the scattering rates. This material is available free of charge via the internet in the arXiv submission .zip file.

\begin{acknowledgement}

N.B. acknowledges the ARCHER UK National Supercomputing Service and the UK Materials and Molecular Modelling Hub for computational resources, which are partially funded by EPSRC (EP/P020194/1). N.B. and F.M. also acknowledge the Cirrus UK National Tier-2 HPC Service at EPCC.
S.P. acknowledge support from the European Unions Horizon 2020 Research and Innovation Programme, under the Marie Sk\l{}odowska-Curie Grant Agreement SELPH2D No.~839217.
F.G.'s contribution to this work was supported as part of the Computational Materials Sciences Program funded by the U.S. Department of Energy, Office of Science, Basic Energy Sciences, under Award DE-SC0020129.

\end{acknowledgement}

\bibliography{biblio}

\end{document}